\documentstyle[floats,multicol,aps,epsfig,prl,subfigure]{revtex}
\newlength{\bxwidth}\bxwidth=3.4 truein
\newcommand\om { \omega}

\newlength{\fight} 
\newcommand{\fg}[4]
{\begin{figure}[h]\epsfxsize=#1
\centerline{\epsfbox{#2}}\vskip 0.1truein
\caption{{#3}}\label{#4}\end{figure}}
\newcommand\ltdash{\raise-1.8pt\hbox{$\scriptscriptstyle |$}}
\newcommand \beq  {\begin{equation}}
\newcommand \eeq  {\end{equation}}
\newcommand \bea {\begin{eqnarray} }
\newcommand \eea {\end{eqnarray}}

\newcommand \dg{^{\dagger}}

\newcommand\bk{{\bf k}}

\newcommand\bq{{\bf q}}

\newcommand\tx{\textstyle}

\begin{document}
\draft
%***********    This is for two columns *******************************
\twocolumn[\hsize\textwidth\columnwidth\hsize\csname @twocolumnfalse\endcsname
%********************************
\title{
Scaling approach to itinerant quantum critical points }
\author{  Catherine P\'epin $^1$, J\'erome Rech $^{1,3}$ and Revaz Ramazashvili$^2$}
\address{$^1$ SPhT, L'Orme des Merisiers, CEA-Saclay, 91191 Gif-sur-Yvette, France
}
\address{$^2$ Materials Science Division, Argonne National Laboratory, Argonne,
IL 60439, USA
}
\address{$^3$ Center for Materials Theory, Rutgers University, Piscataway,
NJ 08855, USA }

\maketitle
\date{\today}
\maketitle
%\widetext
\begin{abstract}
Based on phase space arguments, we develop 
a simple approach to metallic quantum critical points,
% , which enables us to derive the critical properties A
designed to study the problem 
without integrating the fermions 
out of the partition function.
The method is applied to the spin-fermion model of a T=0 
ferromagnetic transition.
Stability criteria for the conduction and the spin fluids
are derived by scaling at the tree level.
We conclude that anomalous exponents may be generated 
for the fermion self-energy and the spin-spin correlation 
functions below $d=3$, in spite of the spin fluid being 
above its upper critical dimension.
\end{abstract}
\vskip 0.2 truein
\pacs{78.20.Ls, 47.25.Gz, 76.50+b, 72.15.Gd}
\newpage
% ***********    This is for two columns *******************************
\vskip2pc]
%********************************

One of the explanations advanced for the breakdown of the Fermi
liquid theory in the normal state of high temperature
superconductors is the proximity to a quantum critical point
(QCP), hidden under the superconducting dome.
The nature of this zero temperature transition remains controversial.
However, in several heavy fermion materials, the Fermi liquid state
was experimentally shown to break down near a well characterized $T=0$
antiferromagnetic instability \cite{julian,mathur,schroeder,custers}. 
A detailed renormalization group study of QCPs in itinerant magnets 
was first undertaken by Hertz~\cite{hertz}, and later augmented by 
Millis~\cite{millis}.
The key observation
~\cite{hertz}
was that spin fluctuations relax critically in time, 
with a dynamic exponent $z$ relating the
time and the length scales as $ \tau \sim \xi^z$.
Thus a $d$-dimensional system can be viewed 
as having effective dimensionality $d+z$.
For an antiferromagnetic QCP, one finds $z=2$,
while for a ferromagnetic QCP, $z=3$.
After integrating the fermions out of the partition function,
the authors of~\cite{hertz,millis} argued that, in $d=2$ or $3$
(the cases of interest for heavy fermions as well as high-T$_c$
superconductors), $d+z \geq 4$. Thus the effective Ginzburg-Landau
theory for the spin fluid falls above its upper critical dimension,
and has Gaussian cri\-ti\-cal behaviour.

Recently, the validity of integrating out the fermions has been
questioned~\cite{chubukov,belitz}, since gapless fermions may lead
to singular coefficients in the Ginzburg-Landau expansion.
Another outstanding question is whether the Fermi liquid theory 
may break down in two or three dimensions, i.e. whether the 
quasiparticles may become ill-defined
while the magnetic fluctuations are only innocuously 
critical, being described by a Gaussian theory. 
In this case, the conduction and the magnetic fluids would 
behave as if decoupled, each having its own upper critical 
dimension. 
Finally, integrating out the fermions to describe a QCP 
in a metal is conceptually unsatisfactory, as it greatly
complicates a consistent account of electron transport.

In this paper, we introdice a simple scaling approach,
designed to study a spin-fermion model at a QCP without
integrating the fermions out of the partition function. 
Already at the tree level, it reveals that
the critical behavior 
is controlled by several couplings, rather than by the
single fermion-boson coupling constant $g$, expected 
naively. At a ferromagnetic quantum critical point, 
the coupling $g$ becomes relevant below one spatial 
dimension, analogously to the four-boson coupling 
constant at a ferromagnetic QCP \cite{hertz}.
At the same time, the four-fermion coupling $u_f$, mediated 
by the bosons, controls the breakdown of the Fermi liquid 
theory, and becomes relevant below three spatial dimensions.

We will illustrate the idea on the spin-fermion model
of a ferromagnetic quantum critical point,
comprising three parts: a conduction electron term $S_f$, a boson
term $S_b$, describing the critical magnetic modes, and $S_{i}$, 
representing interaction between the fermions and the bosons:
\bea \label{eqn1} S &  = & S_f + S_b + S_{i} \\
     S_f & = & \int d \om \ d^d \bk \ \psi^{\dg}_{\bk} \
\big( 
i \om
 - \xi_\bk
  \big) \ \psi_\bk \nonumber \\
     S_b & = & \int d \om \ d^d \bq \; {\bar \varphi}_\bq \ \bigg(
\frac{ |\om |}{q 
}
 + q^2 \bigg) 
\varphi_\bq
\nonumber \\
    S_{i} & = & g \int d \om_1 d \om_2   
d^d \bk_1 d^d \bk_2
\left[ \varphi_{\bk_1 - \bk_2} \psi^{\dg}_{\bk_1} \psi_{\bk_2} + h.c. 
\right]
 \nonumber \ . \eea 
Here $\psi$ ($\varphi$) are the fermion (boson) 
fields,
and $\xi_\bk$ is the quasiparticle energy 
counted from the chemical potential. For 
simplicity, we consider a spherical
Fermi surface.

We first perform a Benfatto-Galavotti-Shankar~\cite{shankar,chetan}
renormalization at the tree level, removing the high energy
degrees of freedom, and retaining only their contribution
to the low energy effective action.
At each step we eliminate a shell $\Lambda/s \leq \om \leq
\Lambda$, where $\Lambda$ is the cut-off and $s >1$.
The energies and momenta are then rescaled to restore the original
cut-offs and, lastly, the fields are rescaled to leave the
quadratic part of the action intact.

However, implementation of this program poses two difficulties.
The first one is of geometric origin: the fermion momenta 
are restricted to a thin shell around the Fermi surface, 
while the boson momenta are confined to a sphere.
The second difficulty stems from the different dynamic exponents $z$ 
of the fermion and boson fluids: in the Fermi liquid, $z_f=1$, 
while the magnetic modes are characterized by $z_b=3$.
To perform simultaneous mode elimination for the two
species, we rescale the energies and the momenta as per
\beq
\label{rescaling1}
\om^\prime = s  \om              ; \; 
k^\prime_{\perp} = s^{1/z_f} k_\perp = s k_\perp ; \; 
q^\prime = s^{ 1 / z_b } q = s^{ 1 / 3 } q , 
\eeq
while the fields rescale as per
\beq
\label{rescaling2}
\psi^\prime = s^{-3/2} \psi ; \; \; \; \; 
\varphi^\prime = s^{ - (d+z_b+2)/2 z_b} \varphi
\eeq
where $k_\perp$ is defined by $\xi_\bk=v_F k_\perp$.
\fg{\fight}{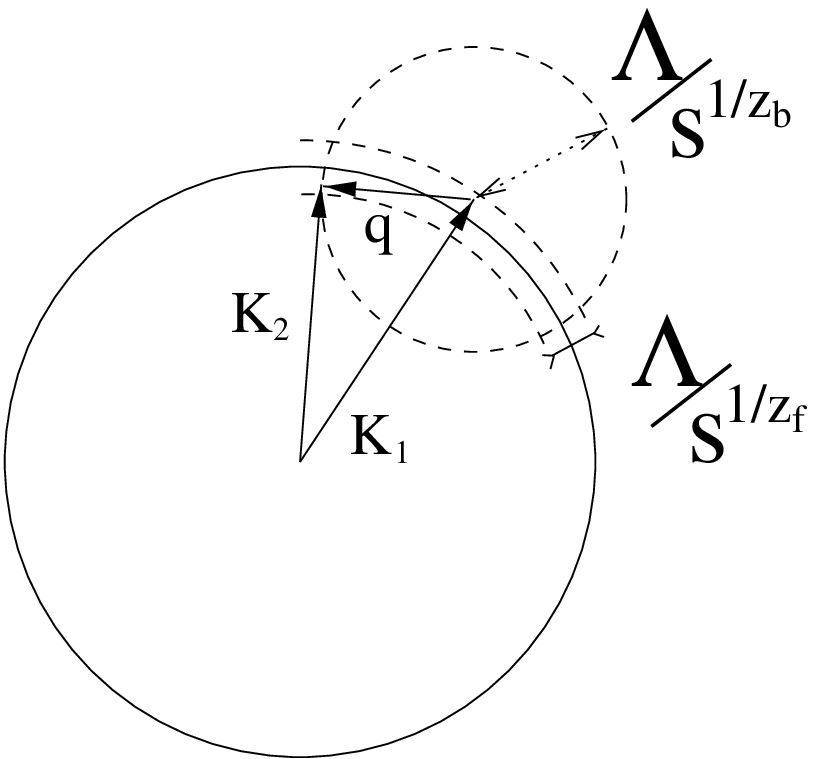}{An illustration of the phase space restriction
for the fermion scattering off magnetic modes, here for a
ferromagnet. The angle between $\bk_2$ and $\bk_1$ is restricted by
the spin fluctuation cut-off that scales as $\Lambda s^{-1/z_b}$.}{figure1}

Before proceeding with the scaling of different quantities,
let us make several observations. 
First, notice an important consequence of the different 
rescaling of boson and fermion momenta: 
since $z_b = 3 > z_f = 1$, 
the cut-off for bosons reduces slower than for fermions
and, near each point at the Fermi surface, the fermion 
scattering processes become restricted to a thin cylindrical slab, 
whose thickness scales as 
$\Lambda s^{-1/z_f} = \Lambda s^{-1}$, 
while its base radius scales as 
$\Lambda s^{-1/z_b} = \Lambda s^{-1/3}$, 
as illustrated in  Fig.~\ref{figure1}.

Hence the fermion momentum transfer occurs 
predominantly parallel to the base of the slab.
As the initial fermion momentum $\bk_1$ 
spans the shell of width $\Lambda$ around the Fermi surface, 
the solid angle of the final fermion momentum $\bk_2$ 
with respect to $\bk_1$ is restricted by the magnetic modes 
(see Fig.~\ref{figure1}).
Thus, to deal with the scattering vertex, 
it is convenient to separate the two fermion
momenta $\bk_1, \bk_2$ into 
$\bk \equiv (\bk_1 + \bk_2)/2$ and   
the transferred boson momentum 
$\bq \equiv \bk_1 - \bk_2$.
Then, we divide both $\bk$ and $\bq$ 
into the one-dimensional components 
$k_\perp$ and $q_\perp$ normal to the Fermi surface, 
and the $(d - 1)$-dimensional components 
$\bk_\parallel$ and $\bq_\parallel$, 
locally tangential to the Fermi surface. 
Note that, according to the above, 
these components scale differently 
under the RG transformation:
\bea
\label{rescaling3}
k^\prime_{\perp} & = & s \ k_\perp 
\ ; \; \; 
d^{d - 1} \bk^\prime_{\parallel}  = d^{d - 1} \bk_\parallel 
\ ; \; \; \\
\label{rescaling4}
q^\prime_\perp & = & s \ q_\perp
\ ; \; \; 
d^{d - 1} \bq^\prime_\parallel = s^{(d - 1) / z_b } \ d^{d - 1} \bq_\parallel
\eea
The physical reason behind $d^{d - 1} \bk_\parallel$ 
not rescaling is that the size of the Fermi 
surface does not change upon rescaling, 
and that $\bk$ spans the entire Fermi surface. 

Finally, let us make an observation regarding 
consistency between our rescaling procedure and momentum 
conservation in a single boson scattering process.
The two fermion momenta
 $\bk_{1, 2} = k_F \hat{k}_{1, 2} (1 + \eta_{1, 2})$, 
where $|\eta_{1, 2}| \ll 1$, are related by 
$\bk_1 - \bk_2 = \bq$. 
Since, near a ferromagnetic QCP, the boson momenta 
are much smaller than $k_F$, 
to first order in $\eta_{1, 2}$ one finds
$$
\theta^2 \left[ 1 + \eta_1 - \eta_2 \right] = 
\left( \frac{\bq}{k_F} \right)^2,
$$
where
 $\cos \theta \equiv (\hat{k}_1 \cdot \hat{k}_2)$, 
and $\theta \ll 1$. 
From this, two points follow: 
(i) rescaling of the boson momentum has to be accompanied
by rescaling of the scattering angle $\theta$ betwen the 
two fermion momenta, as in the procedure we adopted; 
(ii) asymptotycally close to the Fermi surface 
($\eta_{1, 2} \rightarrow 0$), rescaling 
fermion and boson momenta differently 
is consistent with momentum conservation.

Now we are in a position to find the scaling properties 
of various vertices. 
To obtain the rescaling of $g$, rewrite $S_{i}$ as 
$$ 
S_{i} = g \int d\om_1  d \om_2 
d k_\perp d^{d-1} k_\parallel 
d q_\perp d^{d-1} q_\parallel  
( \varphi_\bq
\psi^{\dg}_{\bk_1} \psi_{\bk_2} + h. c. ).
$$  
Using the scaling properties of the fields 
(\ref{rescaling2})
and of the different components of momenta 
(\ref{rescaling3},\ref{rescaling4}),
one arrives at the sought scaling relation: 
\beq \label{rescaling-g}
g^\prime = s^{\tx- \frac{d + z_b - 4}{2 z_b}
} \ g \ .
\eeq
We find that $g$ is irrelevant for $d > 1$, similarly 
to the four-boson coupling constant in the $\varphi^4$ 
theory ($z_b = 3$). \cite{hertz,millis}

% The boson self-energy in the lowest 
% order in $g$ can be evaluated similarly.
% Diagram b) in figure~\ref{figure2} shows that
% $$ 
% \Pi^{(1)} (\bq, \om) \simeq
% g^2 \int d \nu  d k_\perp d^{d-1} k_\parallel 
% G_{\psi} (\nu; \bk) G_{\psi} (\om - \nu; \bf{p} - \bk).
% $$
% Again, the scaling of $d^d k$ is restricted by the vertex
% phase space, and we find
% \beq
% \label{pione} 
% \Pi^{(1)}(s \om, s^{1/3} \bq) = 
% \Pi^{(1)}(\om, \bq)
% s^{ -(d - 1) / z_b} \
% . \eeq
% This corresponds to the starting exponents of the model. 

\fg{\bxwidth}{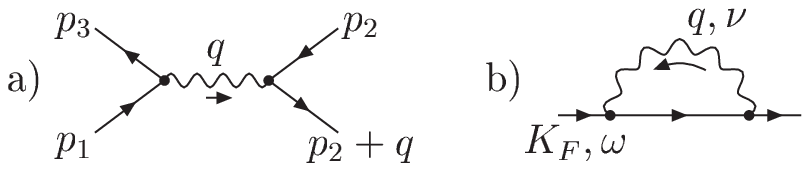}{
%Feynman diagrams associated with a) the
%fermi self-energy; b) and c) the $n$-loop ladder renormalization
%of the boson propagator; d), e) and f) the renormalization of the
%interactions. Solid lines are fermion and wavy lines are boson
%propagators.
Feynman diagrams associated with 
a) the fermi self-energy and 
b) the boson-mediated four-fermi interaction. 
Solid lines are fermion and wavy lines are boson
propagators.
}{figure2}

One may inquire about the relation 
between the coupling constant $u_f$ of the boson-mediated 
four-fermion interaction (Fig.~\ref{figure2} a) 
and the spin-fermion scattering vertex $g$. 
Naively, one would expect $u_f$ to scale as $g^2$ and, 
hence, become irrelevant in one spatial dimension or less; 
we will show that this is not the case: at a ferromagnetic 
quantum critical point, $u_f$ becomes relevant already 
in $d < 3$ spatial dimensions, 
possibly leading to an instability 
of the Fermi liquid ground state. Notice that, 
of the three independent momenta 
($\bf{p}_1$, $\bf{p}_2$, $\bf{p}_3$) 
in the diagram a) in Fig.~\ref{figure2}, only 
the momentum transfer $\bq \equiv \bf{p}_1 - \bf{p}_3$ 
is subject to the phase space restriction discussed 
above, whereas 
$\bf{p} \equiv (\bf{p}_1 + \bf{p}_3)/2$ and $\bf{p}_2$ 
independently span the entire Fermi surface. 
Hence the four-fermion term in the action can
be re-written to make explicit the scaling 
properties of the various components of 
momenta: 
\bea 
S_{4f} &\rightarrow& u_f 
\int
\frac{d \om_1}{s}
\frac{d \om_2}{s}
\frac{d \om_3}{s}
\frac{d p_\perp}{s} 
d^{d-1} {\bf{p}}_\parallel 
\frac{d q_\perp}{s}
\frac{d^{d-1} \bq_\parallel}{s^{(d-1)/z_b}} 
\times \\
 &\times&  
\frac{d {\bf{p}}_{2 \perp}}{s}
d^{d-1} {\bf{p}}_{2 \parallel} \
s^6 
\psi^\dagger_{{\bf{p}}-\bq/2} \ \psi_{{\bf{p}} + \bq/2} \
\psi^\dagger_{{\bf{p}}_2 + \bq} \ \psi_{\bf{p}_2} 
\times \\
 &\times&
\frac{s^{2/z_b}}{\frac{|\om_1 - \om_3|}{|\bq|} + \bq^2},
\eea
from which we read off 
$$
u_f \ \rightarrow \ u_f s^{(3-d)/z_b}.
$$
Which means that, already at the tree level, $u_f$ 
becomes relevant below three spatial dimensions, 
while $g$ is relevant only below $d=1$. 
This indicates that the Fermi liquid state may break 
down below three dimensions, where the naive $\varphi^4$ 
theory would be still above its upper critical
dimension. 

Finally, let us illustrate how one can use 
(\ref{rescaling3},\ref{rescaling4}) 
to find the scaling of the fermion self-energy 
$\Sigma (\om)$ in the lowest order in $g$.
It can be found by power counting of the diagram b) 
in figure~\ref{figure2}; 
its contribution to the self energy is 
$$
\Sigma (\om) \simeq g^2 \int d \nu d q_\perp 
d^{d-1} \bq_\parallel
G_{\psi}(\omega - \nu; \bf{p} - \bq) \ 
G_{\varphi}(\nu; \bq).
$$
Since $G_{\psi}^{-1}(\om; {\bf{p}}) = i\om - v_F p_\perp $ and 
$G_{\varphi}^{-1} (\om, \bq) = |\om|/\bq + \bq^2$, 
one finds, again using 
(\ref{rescaling2},\ref{rescaling3},\ref{rescaling4}):
\beq \label{selfenergy} 
\Sigma(s \om) = \Sigma(\om) s^{-d/z_b}
\eeq
This is in agreement with previous
work~\cite{chubukov,belitz,gan}, and 
points to a possible non-Fermi behavior in $d \leq 3$,
as previously observed in the context of gauge theories~\cite{gan}
and recently noted in the context of a 
ferromagnet~\cite{belitzferro}.

To summarize, we introduced an RG scheme in the spirit of the
Shankar approach~\cite{shankar}, which allows to treat the fermion
and the boson degrees of freedom in the spin-fermion models on an
equal footing.
We showed that, already at the tree level, 
the boson-mediated four-fermion coupling
is relevant below three spatial dimensions, 
even though the fermion-boson coupling constant $g$ 
is relevant only above one dimension.
Our approach is general and can be applied to the theory of 
an antiferromagnetic QCP -- as well as to other situations
where fermions interact with critical modes, e.g. those involving
different dimensionalities of fermions and spin fluctuations. 
The phase space restriction associated with the fermion 
scattering off the critical modes is the key ingredient
of the approach.
A one loop RG treatment in $d = 3 - \epsilon$, including transport properties
and thermodynamics, would be a natural extension of this work~\cite{futur}.

We thank A. Chubukov, P. Coleman, M. Norman, O. Parcollet and J.
Zinn-Justin for very useful discussions and illuminating insights.
RR is supported by the US Dept. of Energy, Office of Science,
under contract No. W-31-109-ENG-38, and would like to thank the 
Aspen Center for Physics for the hospitality during the Summer
of 2003. CP would like to acknowledge the hospitality of Argonne 
National Lab and Rutgers University, where part of the work was performed.

\end{document}